\documentclass[a4paper]{jpconf}
\usepackage{graphicx}
\usepackage{isotope}
\usepackage{amsmath, amssymb, amsbsy}
\usepackage{dcolumn}
\newcolumntype{d}[1]{D{.}{\cdot}{#1} }
\usepackage{cite}
\begin{document}
\title{Possibility of a new neutral hypernucleus \boldmath $\isotope[4][\Lambda\Lambda]{n}=(n,n,\Lambda,\Lambda)$\unboldmath\footnote{Talk given by JMR at SOTANCP3, 3rd International Workshop on
“State of the Art in Nuclear Cluster Physics”, Yokohama, Japan, May 26-30, 2014}}
\author{Jean-Marc Richard$^1$, Qian Wang$^2$, and Qiang Zhao$^3$}

\address{%
$^1$Universit\'e de Lyon, Institut de Physique Nucl\'eaire, UCBL--IN2P3-CNRS,\\
4, rue Enrico Fermi, Villeurbanne, France\\
$^2$Institut f\"{u}r Kernphysik, Institute for Advanced
Simulation,\\ and J\"ulich Center for Hadron
Physics, D-52425 J\"{u}lich, Germany\\
$^3$Institute of High Energy Physics and Theoretical Physics Center for Science Facilities,\\
Chinese Academy of Sciences, Beijing 100049, China}
\ead{j-m.richard@ipnl.in2p3.fr, q.wang@fz-juelich.de, zhaoq@ihep.ac.cn}

\begin{abstract}
The status of the light hypernuclei is reviewed, and discussed with models based either on the Nijmegen-RIKEN  baryon--baryon interaction or on recent studies using chiral effective field theory. The latter suggests a significantly shorter range for the $\Lambda\Lambda$ interaction, and this favours the formation of a Borromean state made of two neutrons and two hyperons. 
Various corrections are discussed, in particular the coupling of $N\Lambda$ to $N\Sigma$, or of $\Lambda\Lambda$ to $N\Xi$, and the role of tensor forces.
The new nucleus $\isotope[4][\Lambda\Lambda]{n}=(n,n,\Lambda,\Lambda)$ could be produced in various reactions, in particular deuteron--deuteron scattering with the simultaneous production of two charged kaons, for which an estimate of the cross section is provided. 
\end{abstract}

\section{Introduction}
The physics of light hypernuclei has been the subject of many papers, and there is a renewed interest in recent months, see, for instance, \cite{Hiyama:2014cua,PhysRevC.89.057001,Gal:2014efa} and references therein. In this contribution, we wish to stress the possibility of Borromean  states such as the neutral $T=(n,n,\Lambda,\Lambda)$ with strangeness $-2$, which is more likely bound using the recent interaction models inspired by chiral effective field theory (CEFT) than in more conventional models based on meson exchange and SU(3) flavour symmetry. 

After a brief reminder on the allowed window for Borromean binding, we shall discuss the interaction models, present our results on light hypernuclei, and then suggest some reaction mechanisms to produce light hypernuclei.

\section{Borromean binding}
By  a kind of miracle, \isotope[6]{He}, considered as $(\alpha, n,n)$ is stable, though neither $\isotope[5]{He}=(\alpha,n)$ nor $(n,n)$ are bound. In quantum mechanics, this can be understood even without invoking 3-body forces. Binding two bosons, for instance, with a short-range potential $g\,V(r)$, where $V(r)$ is attractive and $g>0$, requires a minimal strength, say $g>g_2$. There is also a threshold coupling $g>g_3$ for binding three bosons with the same interaction, $g\,\sum_{i<j} V(r_{ij})$, but the interesting observation is that  $g_3<g_2$! Hence in the window $g\in (g_3,g_2)$, one gets 3-body binding without 2-body binding. This is implicit in the pioneering paper by Thomas \cite{Thomas:1935zz}, who observed a large ratio of 3-body to 2-body binding energies for $g\gtrsim g_2$, and deduced a limit $m_\pi\lesssim 200\,$MeV on the range of nuclear forces before the discovery of the pion. The word ``Borromean'' became popular in the context of halo nuclei \cite{Zhukov:1993aw} and is now of current 
use. 

Borromean binding is allowed in a limited range of couplings. For $n$ bosons, one can demonstrate \cite{Richard:1994mc} that the threshold coupling $g_n$ obey
\begin{equation}\label{eq:gn+1gn}
 (n+1)\,g_{n+1}\ge n\,g_n~.
\end{equation}

A system of unequal masses $(m,m,M)$ involves two couplings $g_{mm}$ and $g_{mM}$, which can be normalised so that $g=1$ corresponds to the coupling threshold for the $(m,m)$ and $(m,M)$ pairs, respectively, one can draw a map of the $\{g_{mm}, g_{mM}\}$ region for which Borromean binding is not forbidden \cite{Richard:1994mc}.
One gets typically one third of the square $\{g_{mm}<1,\, g_{mM}<1$ allowed, so it is conceivable to get $(n,n,\Lambda)$ bound in some models where neither $(n,n)$ nor $(n,\Lambda)$ are stable. 

%This is done in Fig.~\ref{fig:fig1}. 

In the case of a  4-body system with two different masses, i.e. $(m,m,M,M)$, there are three different couplings, $g_{mm}$, $g_{MM}$ and $g_{mM}$. Again, they can be normalised such that $g_{ij}=1$ which is the coupling threshold for the $(m_i,m_j)$ pair. In \cite{Richard:1994mc} and our recent study \cite{Richard:2014pwa}, the best frontier of the allowed region for binding inside the cube $\{g_{mm}<1, \,g_{MM}<1,\, g_{mM}<1$\} is drawn. A large fraction of the cube $\{g_{ij}\le 1\}$ can be eligible for Borromean binding.

This preliminary study provides some  hopes for Borromean binding of $\isotope[4][\Lambda\Lambda]{n}=(n,n,\Lambda,\Lambda)$. Atomic physics gives us some further encouragement to search for Borromean 4-body systems. With a pure Coulomb interaction, the system of four unit charges $(m^+,m^-,M^+,M^-)$ remains bound against dissociation into two neutral atoms in the region of mass ratio $M/m\simeq 2$ (or, of course, $m/M\simeq 2)$, although none of 3-body subsystems such as $(m^+,M^+,M^-)$ or $(m^+,m^-,M^+)$ are stable \cite{PhysRevA.67.034702}.  A group of Como \cite{PhysRevA.69.042504} has studied how the hydrogen atom, ions and molecules evolve when the Coulomb interaction becomes screened: there is a domain of the screening parameter for which the hydrogen molecule remains stable while the hydrogen atom is unbound. 

The inequality \eqref{eq:gn+1gn}, and the allowed domains in the case of unequal masses, are derived from a  Hall--Post type of decomposition of the $n$-body Hamiltonian in terms of $n-1$-body or $2$-body sub-systems, with proper removal of the centre-of-mass motion \cite{Richard:1994mc}. Hence, they are saturated  in the case of a potential with an external barrier and an inner part that coincides with a harmonic oscillator. The chances of getting Borromean binding decrease if the shape of the potential evolves towards a monotonic potential, and further decrease if the potential acquires an inner repulsive core of increasing size and strength \cite{Moszkowski:2000ni}. 

In Refs.~\cite{Richard:2014pwa,PhysRevA.78.020501}, another point of view was adopted, not based on the shape of potential, but on its low-energy scattering properties.   It was shown that for a given scattering length (negative so that the 2-boson system is attractive without bound state), the Borromean binding in the three-boson system depends dramatically on the effective range $r_{\rm eff}$. Moreover, the curve giving the 3-body energy $E_3$ vs.\ $r_{\rm eff}$ is almost universal, i.e., independent of the shape of the potential.

Thus, for our purpose, it is extremely important to use toy potentials that fit accurately not only the scattering lengths but also the effective ranges.

\section{The 3- and 4-body problem}
Our study is devoted to light hypernuclei with mass number $A\le 4$. The 2-body problem has been solved by standard techniques to fix the parameters of some simple potentials as to reproduce the deuteron binding energy, the scattering length and effective range of $\Lambda N$ and $\Lambda\Lambda$ provided by the Nijmegen group and the Bonn-J\"ulich EFT group. 

The 3-body problem has been solved in perimetric coordinates $x_1=|\vec r_2-\vec r_3|$, etc., using the variational wave function
\begin{equation}\label{eq:3bwf}
 \Psi=\sum_i \alpha_i\,\exp[-a_i\,x_1-b_i\,x_2-c_i\,x_3]~,
\end{equation}
that is widely used in atomic physics. The permutation symmetry, if any, can be checked \textsl{a posteriori} or imposed \textsl{a priori} assigning the same coefficient $\alpha_i$ to several terms.  To avoid instabilities in the variational optimisation,  it is convenient to choose the inverse ranges  $a_i$, $b_i$ and $c_i$ in a geometric series, and thus to minimise only two parameters, given that the $\alpha_i$ and the variational energy are computed from a simple eigenvalue equation. This is somehow similar to the method used by Hiyama et al.~\cite{Hiyama:2003cu} for the Gaussian expansion. 

As a check, we also solve the 3-body problem by the method of Gaussian expansion in the stochastic variant of Suzuki and Varga, using their computer code \cite{Varga:1996zz,Varga:1997xga}. This latter method was also applied to the 4-body systems. 
\section{Results}
Once again, our preliminary study is based on a kind of ``model of model'', namely a simple potential that reproduce the low-energy properties of a more complicated model of the baryon-baryon interaction. A similar approach was used in Refs.~\cite{Filikhin:2002wp,Nemura:2002hv}. As a check, we reproduce the results of this latter group, but we note that the models they used, borrowed from the former reference, differ from ours. 

More precisely, we adopted two sets of parameters, given in Table~\ref{tab:para},  the first one,  labelled ESC08, corresponding to the Nijmegen-RIKEN group \cite{Rijken:2013wxa}, the second one, labelled CEFT, to Refs.~\cite{Polinder:2007mp,Haidenbauer:2013oca}. We also use standard values for $np$ in both spin $s=0$ and $s=1$ states (not shown). 
\begin{table}[!htbc]
\label{tab:para}\caption{Low-energy scattering parameters in the two models (in fm)}
\centering
\vskip .1cm
\begin{tabular}{cd{1}d{1}d{1}d{1}}
\hline\\[-10pt]
 & \multicolumn{2}{c}{ESC08} & \multicolumn{2}{c}{CEFT}\\[-2pt]
Pair & a &  r_\text{reff} &  a & r_\text{eff}\\
\hline\\[-5pt]
$nn$ &-16.51 & 2.85 &-18.9 & 2.75\\
$(n\Lambda)_{s=0}$ &-2.7 &2.97 & -2.9&2.65\\
$(n\Lambda)_{s=1}$ &-1.65& 3.63& -1.51 &2.64\\
$\Lambda\Lambda$ & -0.88 &4.34 &-1.54&0.31\\
\hline
\end{tabular}
\end{table}
The most remarkable feature is the much shorter effective range for $\Lambda\Lambda$ in the new CEFT model, as compared to ESC08.

  The results are the following:\\
  $\mathbf{A=2}:$ The deuteron $d=(n,p)$ is of course well reproduced once the low-energy parameters of $np$ are fitted. There are also other bound states with strangeness $S=-1$ or $-2$.\\
   $\mathbf{A=3}:$ The $S= -1$ system $(n,p,\Lambda)$ is bound in both models for isospin $I=0$ and spin $s=1/2$, below the $d+\Lambda$ threshold. With CEFT, the partner with $I=0$ and $s=3/2$ and the one with $I=1$ and $s=1/2$ are nearly bound. We do not find any binding for $(n,n,\Lambda)$ nor for $(n,\Lambda,\Lambda)$, but we have not yet introduced any mixing with configurations involving $\Sigma$ or $\Xi$.\\
   $\mathbf{A=4}:$ We agree with the conclusions of Ref.\cite{Nemura:2002hv} regarding $\isotope[4][\Lambda\Lambda]{H}=(n,p,\Lambda,\Lambda)$.  Our main concern is $\isotope[4][\Lambda\Lambda]{n}=(n,n,\Lambda,\Lambda)$: this state is found unbound with ESC08, in agreement with \cite{Lekala:2014qza}. However, \isotope[4][\Lambda\Lambda]{n} is at the edge of binding with CEFT. For instance, if we simulate the low-energy parameters of the CEFT column in Table~\ref{tab:para} by an exponential potential $V(r)=-g\,\exp(\mu\,r)$ with $g$ and $ \mu$ properly tuned for each baryon-baryon pair in each spin state, we obtain a binding of about $1\,$MeV for $\isotope[4][\Lambda\Lambda]{n}$ using about 80 terms (plus the ones deduced by permutation) in the Gaussian expansion scheme~\cite{Varga:1996zz,Varga:1997xga}.

Several improvements might enforce or spoil the binding. In particular it would be desirable to push further the CEFT calculations to test the stability in the 2-body sector and access to the 3-body forces. It is well-known that in the strangeness $S=0$ sector, 3-body forces are crucial to reproduce the properties of light nuclei.

Another concern is the neglect of tensor forces, which, in our approach, are replaced by a reinforcement of the effective central interaction. The validity of this approach was discussed many years ago and summarised in the textbook by Blatt and Weisskopf \cite{Blatt:628052}. It is argued that in the deuteron, the tensor forces couple D-states in the continuum to a system located at $-2\,$MeV, while in the triton, they couple the same continuum to a system lying at $-8\,$MeV. This explains while models with pure central forces simulate too much tensor interaction beyond the deuteron, and in particular, overbind the triton. In the case of, e.g., the $N\Lambda$ interaction, the zero-energy where the scattering length and the effective range are fixed almost coincides with the energy of the very weakly bound states such as $(n,n,\Lambda,\Lambda)$, and the tensor forces playing a comparable role in 2-body and few-body systems, can be mimicked more safely within  the same central potential. 

A related important issue is the role of the couplings $\Lambda N\leftrightarrow \Sigma N$ and $\Lambda\Lambda\leftrightarrow \Xi N$. Part of it can be absorbed into an effective diagonal potential in the $\Lambda N$ and $\Lambda\Lambda$ channels. However, as stressed in the literature \cite{Hiyama:2014cua,PhysRevC.89.057001,Gal:2014efa,Gibson:1977ch,Garcilazo:2014pxa}, the $\Sigma N$ and $\Xi N$ channels sometimes open new isospin  couplings within the hypernuclei, and this will lower the energy.

The shape and the strength of the $\Lambda\Lambda$ is intimately related to the existence of the $H$ dibaryon. The chromomagnetic binding of $H=(u,u,d,d,s,s)$ below the $\Lambda\Lambda$ threshold was proposed by Jaffe \cite{Jaffe:1976yi}, with, however, some unjustified assumptions about the SU(3) flavour symmetry and the short-range correlations in the 6-body system \cite{Oka:1983ku}. Many experiments have searched for the $H$, without success. It remains that the chromomagnetic interaction at the quark level induces a short-range attraction between two $\Lambda\Lambda$, and it is interesting that this property is recovered in the CEFT approach. If the $\Lambda\Lambda$ system itself is not bound, the attraction might be sufficient to bind \isotope[4][\Lambda\Lambda]{n}.
\section{Production of \boldmath\isotope[4][\Lambda\Lambda]{n} \unboldmath}
Several reactions have been used to produce double hypernuclei. One possibility is to use a $\Xi^-$ exotic atom as initial state \cite{Zhu:1991zq}. Another way is to consist of $(K^-,K^+)$ inelastic scattering on a light nucleus, with a recoiling system of strangeness $S=-2$ \cite{Yamamoto:1997pa}. There is also the possibility of inclusive production in electron or hadron scattering. 

In the case of the tetrabaryon $T=\isotope[4][\Lambda\Lambda]{n}$, the exclusive reaction 
\begin{equation}\label{eq:reac}
d+d\to T+K^++K^+~, 
\end{equation}
where $d$ is the deuteron, 
seems particularly promising \cite{Richard:2014pwa}. The leading mechanism is a virtual $\pi^0$ exchanged between the two protons. This means twice the reaction $\pi^0+p\to K^+\Lambda$, whose cross-section is known. One can thus estimate rather safely the order of magnitude of the reaction \eqref{eq:reac}, given the very low background. 
Our estimate \cite{Richard:2014pwa} gives a cross-section of the order of $1-2\,$nb for an energy $\sqrt{s}\sim 5.5\,$GeV. Thus the reaction  \eqref{eq:reac} appears as an interesting tool to search for new hypernuclei.

\section{Outlook}
The recent studies of the hyperon-nucleon and hyperon-hyperon interaction indicates a significantly shorter range, as compared to the conventional approaches. This modifies the spectroscopy of light hypernuclei, and conversely, the study of light hypernuclei can probe the new models of baryon--baryon interaction.

We presented here the first results of a study restricted to simple pairwise potentials that reproduce the low-energy properties of each baryon-baryon pair. 
In the case where the potentials are tuned to reproduced the low-energy scattering parameters of Chiral Effective Field Theory, there are indications that the tetrabaryon $\isotope[4][\Lambda\Lambda]{n}=(n,n,\Lambda,\Lambda)$ might be bound. If so, this would be a fully Borromean system whose 2-body as well as 3-body sub-systems are unbound. 
The role of 3-body forces and the changes induced by the coupling of $\Lambda N$ to $\Sigma N$ and of $\Lambda\Lambda$ to $\Xi N$ will be the subject of further investigations.

\subsection*{Acknowledgments}
We thank the organisers for this very stimulating meeting SOTANCP3, providing the opportunity to discuss with many experts.The participation by JMR was made possible by the French-Japanese project CNRS-JPS \#169725. We are also grateful to Prs. V.~Belyaev and A.~Gal for useful comments.

This work was done within the China--France program FCPPL, and  is also supported, in part,
by the National Natural Science Foundation of China (Grant No.
11035006), the Chinese Academy of Sciences
(KJCX3-SYW-N2), and DFG and NSFC funds to the Sino-German CRC 110. 

\section*{References}
%  \bibliography{hypernuclear}
%  \bibliographystyle{iopart-num}
%  \end{document}

\providecommand{\newblock}{}

\end{document}